\documentclass[conference]{IEEEtran}  % Comment this line out
\IEEEoverridecommandlockouts                              % This command is only
\usepackage{graphicx}
\usepackage{subcaption}
\usepackage{multirow}
\usepackage{tikz}
\usepackage[fleqn]{amsmath}
\usepackage{pstricks}
\usepackage[nottoc]{tocbibind}
\usepackage{tikz}
\usepackage{verbatim} 
\usetikzlibrary{backgrounds}
\tikzset{use path/.code=\tikz@addmode{\pgfsyssoftpath@setcurrentpath#1}}
\usetikzlibrary{shapes,snakes}
\usetikzlibrary{positioning,arrows}
\usetikzlibrary{patterns}
\usetikzlibrary{fit}
\usetikzlibrary{graphs}
\usepackage[fleqn]{amsmath}
\usepackage[figuresright]{rotating}
\usepackage{graphicx}
\usepackage{amssymb}
\usepackage{lineno}
\usepackage{multirow}
\usepackage{colortbl}
\usepackage{caption}
\usepackage{marvosym}
\usepackage{cite}
\usepackage{pgf}
\usetikzlibrary{arrows,automata}
\usepackage[latin1]{inputenc}
\graphicspath{ {images/} }
\usepackage[english]{babel}
\makeatletter
\newcommand*\titleheader[1]{\gdef\@titleheader{#1}}
\AtBeginDocument{%
  \let\st@red@title\@title
  \def\@title{%
    \bgroup\normalfont\large\centering\@titleheader\par\egroup
    \vskip1.5em\st@red@title}
}
\makeatother
\title{\LARGE \bf Contract-connection:An efficient communication protocol for Distributed Ledger Technology
}
\titleheader{\small 2019 IEEE 38th International Performance Computing and Communications Conference (IPCCC) 10.1109/IPCCC47392.2019.8958730}

\author{\centering
\IEEEauthorblockN{1\textsuperscript{st} Yibin Xu}
\IEEEauthorblockA{\textit{School of Computer Science and Informatics} \\
\textit{Cardiff University}\\
Cardiff, UK \\
work@xuyibin.top}
\and
\IEEEauthorblockN{2\textsuperscript{nd} Yangyu Huang}
\IEEEauthorblockA{\textit{School of Electronic Engineering and Automation} \\
\textit{Guilin University of Electronic Technology}\\
Guilin, China \\
i@hyy0591.me}
}

\begin{document}

\maketitle
\thispagestyle{empty}
\pagestyle{empty}

%%%%%%%%%%%%%%%%%%%%%%%%%%%%%%%%%%%%%%%%%%%%%%%%%%%%%%%%%%%%%%%%%%%%%%%%%%%%%%%%
\begin{abstract}
Distributed Ledger Technology (DLT) is promising to become the foundation of many decentralised systems. However, the unbalanced and unregulated network layout contributes to the inefficiency of DLT especially in Internet of Things (IoT) environments, where nodes connect to only a limited number of peers. The data communication speed globally is unbalanced and does not live up to the constraints of efficient real-time distributed systems. In this paper, we introduce a new communication protocol, which enables nodes to calculate the tradeoff between connecting/disconnecting a peer in a completely decentralised manner. The network layout globally is continuously re-balancing and optimising along with nodes adjusting their peers. This communication protocol weakened the inequality of the communication network. The experiment suggests this communication protocol is stable and efficient.
\end{abstract}
\section{INTRODUCTION}
Distributed systems, where computational entities are connected to and organised by networks to work collectively in large-scale and high performance, have earned significant attention in contemporary life \cite{jiang2012rich,jiang2011decision,ye2013self}. Distributed Ledger Technology (DLT) is one kind of decentralised system that of replicated, shared, and synchronised digital data geographically spread across multiple sites, countries, or institutions \cite{walport2016distributed}. The first well knew DLT---Nakamoto blockchain, and most permitless blockchains \cite{cachin2016architecture,kiayias2017ouroboros} require participants to accept the first valid block (statement) posted by one participant in every fixed period (referred to as the block interval), the block is built on top of the previous accepted block. Thus, a balanced network structure of the communication protocol running below these DLTs is vital for the fairness of the system as the earlier a node finished hearing a block, the earlier it starts to create the next block. A faster or slower sub-network will slow down the network in overall, DLT must has an extended block interval to enable most nodes in heterogeneous network environments to sync data and to create blocks. Sadly, DLTs like Bitcoin \cite{nakamoto2008bitcoin} suffers from a slow and unbalanced network. It is observed that blocks first propagated by the fastest node reach $50\%$ of the nodes in $2.3s$ whereas blocks first propagated by the slowest node reach $50\%$ of the nodes in more than $1,800s$ with merely over $6,000$ nodes in 2016 \cite{scardovi2016restructuring}. Given the decentralised and distributed nature, how the entities inside the DLT network collaborate to balance the network structure and improve efficiency is a severe problem. 

Yet, the study toward optimising the communication protocol of DLT has not been placed similar attention as like the attention for extending the block throughput or transaction per second \cite{kokoris2018omniledger,zamani2018rapidchain,eyal2016bitcoin}. Though many may argue that the redundancy of the network structure is beneficial for fault tolerance because it is common for nodes to go offline without prior notice, network readdresses in the current protocol is seldom needed. The tradeoff between fault tolerance and the speed of data propagation as well as the fairness of the system is worth studying.

In this paper, we discuss a new communication protocol for DLT, which achieves an equilibrium network structure through a connection adjustment method. This connection adjustment method is of local optimisation (accelerate the speed for hearing the data propagated from any direction in the network) for global optimisation (make the speed for data propagation started from the most point in the system reached the majority of nodes at a similar time).

\section{Hypothesis and Approach overview}
\subsection{Hypothesis}
Any node in the system can publish a transaction, block or gossip message. Thus, every node gets a chance to be closer to a data publisher if they are directly connected or can reach each other within some levels of the network. When the system is data-extensive, and the network structure is well organised, a node which has a higher number of links should be able to hear more data faster in a fixed time window provided there are few redundancy connections. Thus, it should be able to re-transmit more data to its peers faster. If there are three nodes (node A, node B and node C), C peered both A and B; A and B have a similar number of connections and A has a faster communication speed to node C. Then C should be able to hear more pieces of data sooner from A in a fixed time window. If the nodes of similar conditions are categorised into groups, the lazy nodes can be filtered out, and nodes gained the ability to judge others. To not be determined by others as a lazy node, a node should continuously evaluate the performance of its peers and adjust its peers. In this way, every node is seeking to optimise its peers while locating itself to the best position in the network structure.

The challenges of this hypothesis are (1) how to acquire the accurate connection number of every node? (2) when there are only minimal data flowed in a period, how to make the performance measurement? (3) how to quantify the performance of the nodes and how to derive a standard performance for nodes of similar performance? (4) how to categorise nodes and how to avoid nodes peer too many or too little nodes? (5) which peers should a node connect? (6) when should a node replace bad performance peers? 

\subsection{Approach}
\subsubsection{Publish connections to blocks} {When building a connection, each side of the connection co-sign a statement (referred to as a contract) and send this contract to the blockchain. The contract contains the identity as well as the IP and Port of each side of the connection. When terminating a connection, either side of the connection should publish this information to the blockchain.}
\subsubsection{Peering restriction and peer score}{
We rule that two nodes can become peers only when they don't have a mutual peer. This design motivates peers to consider the tradeoff between building a connection with another node and the restrictions of peering after this connection is created. For every node, every peer of it is being marked by a combined index of the communication speed, the structure of this peer's peers and the number of peers this peer has. The performance of the peers of a similar score is compared by the number of data pieces this node first heard from them in a fixed time window.}
\subsubsection{Send data in pieces} {For any data larger than $500bytes$, it is split into parts with each portion sized $500bytes$ in maximum. The data publisher should not send all the portions to a peer and then move on to the next peer; instead, it should send different parts of data to different peers per time until it shipped all the parts of data to all of its peers. This method accelerates the data transmit as it is not necessary for nodes to finish hearing a data before re-transmitting the data. Assumed a IoT device $Gary$ linked itself to ten other IoT devices and all these devices are inside the same category, then $Gary$ should hear approximately the same number of pieces/transactions from every device; if the devices are in different categories, then the number of data pieces should be within the corresponding ranges.}
\subsubsection{Evaluate the performance of nodes}{
If a peer showed an abnormal performance among the peers of the similar peer score, the node might disconnect this peer. Nodes are motivated to evaluate their peers because they want honest and diligent peers to accelerate the speed of hearing overall. If they don't control their peer qualities, they may be considered as abnormal for others. For example, because the data propagation in DLT relies on voluntary re-transmissions, if some of a node's peers are not re-transmitting an adequate number of data to this node that fitted their peer scores in a fixed time window, this node will then have less data that can be re-transmitted to its other peers. When a peer compares this node's performance with this peer's other peers of a similar score, this node may be considered of low performance because it transmitted an unfulfilled number of data pieces.}
\subsubsection{Autonomous decision on peering/unpeering}{
By evaluating the performance of a peer in a fixed time window, a quantified performance score can be calculated. The number of peers, the average performance of all the peers as well as the average scores of peers are used to feed the reinforcement learning (RL) algorithm \cite{van2016deep}. The algorithm can decide to add peers or to replace peers or to do nothing at the end of every time window. The differences in the average time for receiving all the parts of data for every data iteration happened in this time window and that in the last time window are the reward for the decision made by the RL algorithm.}
\section{CONTRACT-CONNECTION PROTOCOL}
\subsection{Definitions}
\begin{itemize}
    \item Data propagation. {When a node broadcast data to the network. If the data is larger than 500 bytes, it is divided into parts; a part is sized 500 bytes in maximum. Before a data propagation begins, a data header of a tiny size (34 bytes) is sent to the network, which indicates the type of data (e,g. blocks, transactions) and the Merkle Root of the data. Then the node sends the divided parts to its peers. Different parts are being sent to different peers at the same moment. When all the parts are being sent out, this marks the end of a moment; then the next moment starts until every peer has heard the entire data for the data sender. This procedure is showed in Figure \ref{fig:img2}, in which the node has four peers $1$, $2$, $3$ and $4$; data is split into three parts.
    \begin{figure}
    \centering
        \begin{tikzpicture}[-,>=stealth',shorten >=1pt,auto,node distance=1.2cm,
                    semithick,scale=0.85,L1Node/.style={circle,   draw=blue!50, fill=blue!20, very thick, minimum size=10mm},
    L2Node/.style={rectangle,draw=green!50,fill=green!20,very thick, minimum size=4mm}]
      \tikzstyle{every state}=[fill=red,draw=none,text=white]

       \foreach \x in {1,...,4}
        \node[state] (w1_\x) at (0,1*\x){\footnotesize Peer$\x$};
        \foreach \x in {1,...,4}
        \node[draw=none,fill=none] (m_\x) at (1.7*\x,5){\footnotesize $Moment \x$};
         \foreach \x in {4,...,1}
        \node[L2Node,draw=gray!50,fill=gray!20] (m1_\x) at (1.8,1*\x){\footnotesize Data header};
         \foreach \x in {1,...,4}
         \draw[->] (m1_\x) --(w1_\x);
         \foreach \x in {1,...,4}
        \node[L2Node,draw=gray!50,fill=gray!20] (m3_2) at (1.7*2,1){\footnotesize Part 1};
        \node[L2Node,draw=gray!50,fill=gray!20] (m3_3) at (1.7*2,2){\footnotesize Part 3};
        \node[L2Node,draw=gray!50,fill=gray!20] (m3_3) at (1.7*2,3){\footnotesize Part 2};
        \node[L2Node,draw=gray!50,fill=gray!20] (m3_3) at (1.7*2,4){\footnotesize Part 1};
        
        \node[L2Node,draw=gray!50,fill=gray!20] (m3_2) at (1.7*3,1){\footnotesize Part 2};
        \node[L2Node,draw=gray!50,fill=gray!20] (m3_3) at (1.7*3,2){\footnotesize Part 1};
        \node[L2Node,draw=gray!50,fill=gray!20] (m3_3) at (1.7*3,3){\footnotesize Part 3};
        \node[L2Node,draw=gray!50,fill=gray!20] (m3_3) at (1.7*3,4){\footnotesize Part 2};

        \node[L2Node,draw=gray!50,fill=gray!20] (m4_2) at (1.7*4,1){\footnotesize Part 3};
        \node[L2Node,draw=gray!50,fill=gray!20] (m4_3) at (1.7*4,2){\footnotesize Part 2};
        \node[L2Node,draw=gray!50,fill=gray!20] (m4_3) at (1.7*4,3){\footnotesize Part 1};
        \node[L2Node,draw=gray!50,fill=gray!20] (m4_3) at (1.7*4,4){\footnotesize Part 3};
    \end{tikzpicture}
    \caption{An example of publishing a data to the network}
    \label{fig:img2}
    \end{figure}
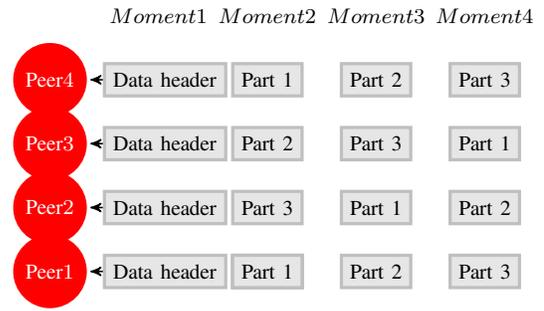}
    \item Peer. {When the contract between two nodes is embedded in the blockchain, the two nodes are peers to each other until the contract is terminated. Two nodes can become peers to each other only when they don't have a mutual peer.}
    \item Peer List (PL). {PL is a set of peer information. For every peer of a node, the peer list records the NID (a 32 bytes public key) as well as the IP, Port of the peer and the co-signed contract. $PL_A$ stands for the PL of node A. Peer Number (PN). $PN_A$ is the number of peers the node A has.
    \item Index of Peer Coincidence (IPC). $IPC_{A,B}=\frac{Card(SubPL_B\backslash PL_A)}{PN_B}$, where $SubPL_B$ is the set of $PL$ of all the peers of node B. Figure \ref{fig:img3} shows an example of the $PL$ and $SubPL$. For this example, $IPC_{A,B}=\frac{6-2}{4}=1$; $IPC_{B,A}=\frac{7-2}{4}=\frac{5}{4}$.
    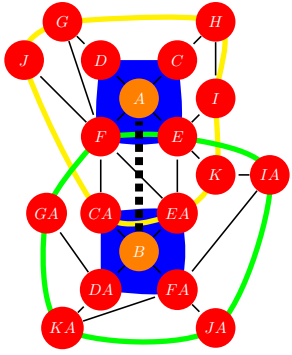
\begin{figure}
        \centering
    \begin{tikzpicture}[-,>=stealth',shorten >=1pt,auto,node distance=1.2cm,
                    semithick,scale=0.5]
  \tikzstyle{every state}=[fill=red,draw=none,text=white,scale=.6]

  \node[state,fill=orange] (A)                    {$A$};
  \node[state]         (C) [above right of=A] {$C$};
  \node[state]         (D) [above left of=A] {$D$};
  \node[state]         (E) [below right of=A] {$E$};
  \node[state]         (F) [below left of=A]       {$F$};
  \node[state]         (G) [above left of=D]       {$G$};
  \node[state]         (H) [ above right of=C]       {$H$};
  \node[state]         (I) [ above right of=E]       {$I$};
  \node[state]         (J) [  below left of=G]       {$J$};
  \node[state]         (K) [  below right of=E]       {$K$};
  \node[state]         (EA) [ below left of=K]       {$EA$};
   \node[state]         (IA) [    right of=K]       {$IA$};

  \node[state,fill=orange]         (B) [ below left of=EA]       {$B$};
  \node[state]         (CA) [ above left of=B]       {$CA$};
  \node[state]         (DA) [ below left of=B]       {$DA$};
\node[state]         (GA) [  left of=CA]       {$GA$};

  \node[state]         (FA) [ below right of=B]       {$FA$};

  \node[state]         (JA) [ below right of=FA]       {$JA$};
  \node[state]         (KA) [ below left of=DA]       {$KA$};

  \begin{scope}[on background layer]
             \fill[blue] plot [smooth] coordinates{ (D) (F) (E) (C)};
      \fill[blue] plot [smooth] coordinates{(CA)  (EA) (FA) (DA)};
     \draw[draw=yellow,line width=2pt] plot [smooth]  coordinates {(G)  (H) (I)  (K) (EA)  (CA) (J) (G) };
          \draw[draw=green, line width=2pt] plot [smooth]  coordinates {(F) (E) (IA) (JA) (KA) (GA) (F)};
    \end{scope}

  \path (A) edge (C)
            edge (D)
            edge (E)
            edge (F);
  \path (G) edge (D)
            edge (F);
  \path (H) edge (C)
            edge (I);
  \path (E) edge (I)
            edge (K);
    \path (F) edge (J);
  \path (IA) edge (FA)
            edge (K);
  \path (EA) edge (E)
            edge(F);
\path (CA) edge (F);
  \path (B) edge (CA)
            edge(EA)
            edge(DA)
            edge(FA);
    \path(FA) edge(JA);
    \path(KA) edge(FA)
              edge(DA);
    \path(DA) edge (GA);
    \path [dotted, draw=orange, line width=3pt](B) edge (A);

    %\filldraw[blue] (D) edge (C) edge (E) edge (F);
\end{tikzpicture}
        \caption{An example of $PL$ and $SubPL$, where $PL_A$ contains four nodes in the blue with node A; $PL_B$ contains four nodes in the blue with node B; $SubPL_{A}$ contains seven nodes linked in the yellow line; $SubPL_{B}$ contains six nodes linked in the green line.}
        \label{fig:img3}
    \end{figure}}
    \item Network Distance (ND). {$ND_{A, B}$ is the Network Distance between node A and node B, which is defined as $\frac{1Mbytes}{Tt}$ where $Tt$ represented the time in second consumed for node A to retrieve a data that sized $1Mbytes$ from Node B.}
    \item Structure Proportion (SP). {$SP_{A,B}=PN_{B}*IPC_{A,B}*(1+ND_{A,B})$.}
    \item Grubbs criterion (X) {is the Grubbs criterion algorithm \cite{grubbs1950sample}, which is used for separate outliers; Grubbs criterion (X) has four steps:
    \begin{enumerate}
        \item If $X=\empty$ or $Card(X)<3$, return $X$.
        \item If $\frac{|\bar{X}-X_1|}{S}>=GrubbsTable(Card(X),p=0.95)$, then $X=X\backslash X_{1}$, repeat (2);
        \item If $\frac{|\bar{X}-X_{Card(X)}|}{S}$ $>=$ $GrubbsTable(Card(X),p=0.95)$, then $X=X\backslash X_{Card(X)}$, repeat (3);
        \item Return $X$.
    \end{enumerate}
    where, $\bar{X}=\frac{X_1+X_2+...+X_{Card(X)}}{Card(X)}$, $S=\sqrt{\frac{\sum _{i=1}^N (X_i-\bar(X))^2}{Card(X)}}$.}
    \item NFHDP. {$NFHDP_A$ is the number of parts of data that is first heard from peer A in a data propagation.}
    \item NFHDC. {$NFHDC_A$ is the number of parts of all the data that is received since the node built a connection with peer $A$.}
    \item ExpNFHDP. {When the node $A$ finished hearing data from one data propagation, 
    \begin{enumerate}
        \item     it creates a set of arrays $ODP$. $ODP_{-\infty...+\infty}=\empty$ is the initial value.
        \item Let $ODP_{SP_{A,i}}=NFHDP_i, i\in[1,PN_A]$. 
        \item The $ExpNFHDP$ for node A's peers are $EXPNFHDP_i=Average(Grubbs\  criterion(\{ODP_{SP_{A,i}-T},$ $...$ $,ODP_{SP_{A,i}+T}\}))$, $i\in[1,PN_A]$. T is a parameter that will be adjusted in the RL algorithm. Figure \ref{fig:img4} shows an example of ExpNFHDP.
    \end{enumerate}
\begin{figure}[h]
    \centering
    \includegraphics[width=0.35\textwidth]{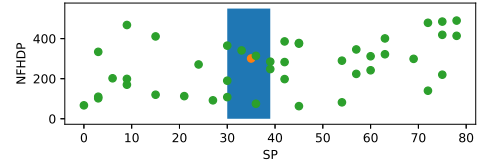}
    \caption{An example of $ExpNFHDP$, the dot in orange represents the $ExpNFHDP$ for $SP_{A,i}=35$ and $T=5$; the green dots in the blue rectangle are the values used in Grubbs criterion.}
    \label{fig:img4}
\end{figure}}
\item ExpNFHDC. {Supporting we are operating on node A. $ExpNFHDC_i$ is the sum of the $ExpNFHDP_i$ during the contract between peer $i$ and node $A$. Every time a data propagated, $ExpNFHDC_i=ExpNFHDP_i+ExpNFHDC_i$, $i \in [1,PN_A]$. $ExpNFHDC_i=0$ is the initial value when node $A$ and node $i$ built a connection.
\item Determine Index (DI). $DI_{A,B}=sin(min(\frac{3}{2}\pi,\frac{NFHDC_B+1}{ExpNFHDC_B+1}*\frac{\pi}{2}))$.
\item Fulfill Rate (FR). $FR_{A,S,E}=sin(min(\frac{3\pi}{2},P*\frac{\pi}{2})),$ $P=\frac{\sum_{j=0}^{PN_A}{\sum_{i=1}^{N_{S,E}}{\frac{NFHDP^i_j+1}{ExpNFHDP^i_j+1}}}}{PN_A}$, where $N_{S,E}$ is the number of data propagation during the time in second $S$ to $E$, $NFHDP^i_j$ and $ExpNFHDP^i_j$ is $NFHDP_j$ and $ExpNFHDP_j$ at $i$ data propagation in this duration respectively.}
\item Average bandwidth (AB). {Let $D1$ be the time when the data header of one data propagation is received; let $D2$ be the time when all the parts of one data propagation is received. $AB_{B,E}=Average(\frac{D2_j-D1_j}{DATA_{SIZE_{j}}}), j \in N_{S,E}$, where $D1_j$ and $D2_j$ are the $D1$ and $D2$ at the number $j$ data propagation during the time in second $S$ to $E$, $DATA_{SIZE_{j}}$ is the data size of $j$ data propagation.}
\end{itemize}
\subsection{Automatic operations}
We use two Q-learning \cite{van2016deep} models to make automatic operations for every node. One (referred to as $Alice$) decides whether the peers of the node should be adjusted; another (referred to as $Bob$) decides whether a node should accept the connection invitation from another node. Let the current time (in second) be $C$, $Alice$ and $Bob$ will be activated every time $C\ mod\ W=0$; where $W$ is a random parameter that is different from nodes to nodes. We set $W \in [30,600]$ assumed the block interval is $30$ seconds.
\subsubsection{Alice} $Alice$ is a tuple.
{
\begin{itemize}
    \item State = ($FR_{X,C-W,C}$, $PN_X$),
    \item Action=\{Add, Replace1, Replace2, STAY, ChangeT1, ChangeT2\},
    \item Reward=$AB_{C-W,C}-AB_{C-2*W,C-W}$,
    \item Policy.
\end{itemize}}
where ADD refers to a function that add a new peer which fulfill the following conditions.
{
\begin{enumerate}
    \item The candidate node accepts to build new connection.
    \item Connect to candidate node will not violate the connection restriction (they don't have a mutual peer).
    \item The candidate node is of the highest $SP$.
\end{enumerate}}

Replace1 refers to a function that delete the peer of smallest $DI$ and ADD a new peer. This action is conducted at the same time. So that the contract with the new peer served as the both termination notice and the new connection contract. Replace2 refers to a function that delete the peer of the highest $IPC$ from the current node's perspective and ADD a new peer; Other operations the same as Replace1. STAY operation refers to a function that doing nothing. ChangeT1 is a function that add value 0.25 to $T$. ChangeT2 is a function that reduce value 0.25 of $T$. ChangeT1 and ChangeT2 can be conducted at the same time with either one of Add, Replace1, Replace2 or STAY operation.
\subsubsection{Bob}
\begin{figure}[h]
    \centering
    \includegraphics[width=0.30\textwidth]{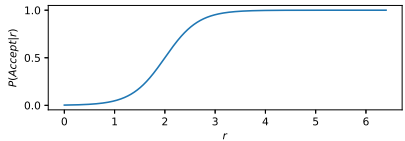}
    \caption{An example of $P(Accept|r)$}
    \label{fig:img5}
\end{figure}

{ $Bob$ is a tuple.
\begin{itemize}
    \item State = ($FR_{X,C-W,C}$, $PN_X$),
    \item Action=\{Allow, Not allow\},
    \item Reward=$-(RJ_{C-W,C}-RJ_{C-2*W,C-W})$,
    \item Policy.
\end{itemize}}
where $RJ_{i,j}$ is the number of connections that were terminated by the other side of the connections during the time between $i$ to $j$. "Allow" refers to the setting that the node will accept the connection invitation from others in the next time window; Not Allow refers to the setting that the node will not accept the connection invitation from others in the next time window. Let $r=\frac{LT-C}{10}$, where $LT$ is the last time when node $i$ built a connection with node $A$; $C$ is the current time. The chance for node $A$ to accept the connection invitation from node $i$ when node $A$ is accepting invitations at the current time window is $P(Accept|r)=\frac{1}{exp(6-3r)}$ provided building this connection is not violating the connection restriction. If node $i$ did not build a connection with node $A$ before then $LT=0$. Figure \ref{fig:img5} shows an example of $P(Accept|r)$.

\section{THE EXPERIMENT}
The purpose of our experiment is to survey Contract-connection protocol performance. We want to show the benefit of the balanced network layout through testing the time for broadcasting the data at random places in the network (the differences in the time between the broadcasting started and finished). We add Bitswap as the comparison.

\begin{figure}[h]
    \centering
    \includegraphics[width=0.35\textwidth]{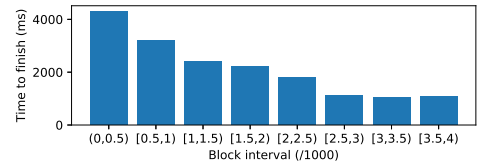}
    \caption{Average time for all the nodes to complete hearing a $1Mbytes$ data, with the progress of block intervals.}
    \label{fig:img7}
\end{figure}

We use two emulated networks, one with $2000$ Nakamoto blockchain nodes which run on contract-connection protocol; another with $2000$ nodes run on Bitswap. To show the comparison, the capacity of nodes in these two networks are mirror images to each other: if there is a node which has full duplex of a specific bandwidth, there will also be one node of the same setting in the other network. Every node of the system is randomly given a fixed upload bandwidth speed ranged from $50Kbytes/s$ to $5Mbytes/s$. When establishing a connection, a network delay time ranged from $10ms$ to $600ms$ is given to this connection. If the delay time of the connection between node $A$ and node $B$ of one network is $60ms$ then the delay time between node $A$ and node $B$ of the other network is also $60ms$. Figure \ref{fig:img6} shows the basic statics of the two networks. We set the block interval time for this experiment to be 30 seconds; every node sends one to three transactions in every iteration of the game. $W$ of every node is set to be a random number between $30$ to $60$. We set up two random connections for every node at the beginning of the game. Figure \ref{fig:img7} shows the average time between a data sized $1Mbytes$ is broadcasted, and it is received by all the nodes with the progress of the game.
\begin{table}[]
    \centering
    \begin{tabular}{ll}
             \includegraphics[width=0.23\textwidth]{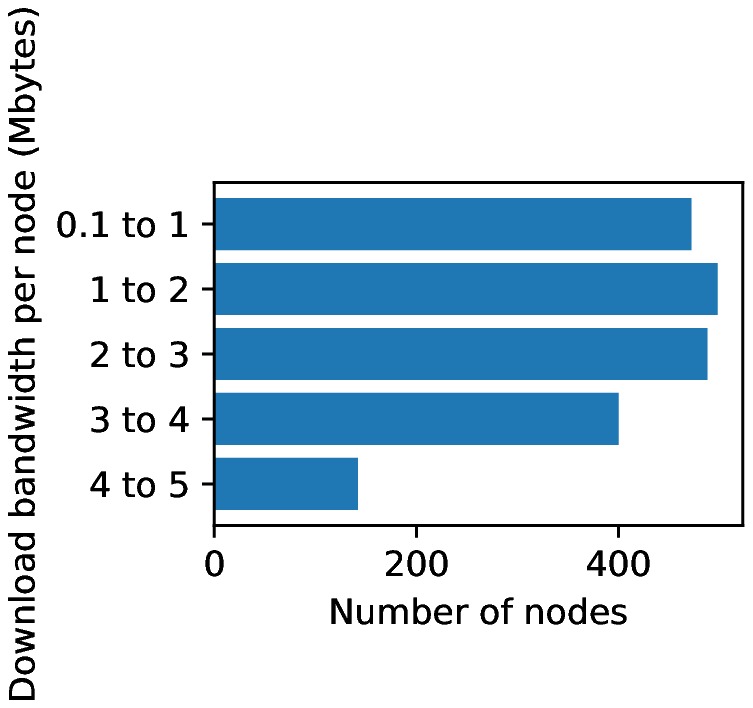}& \includegraphics[width=0.23\textwidth]{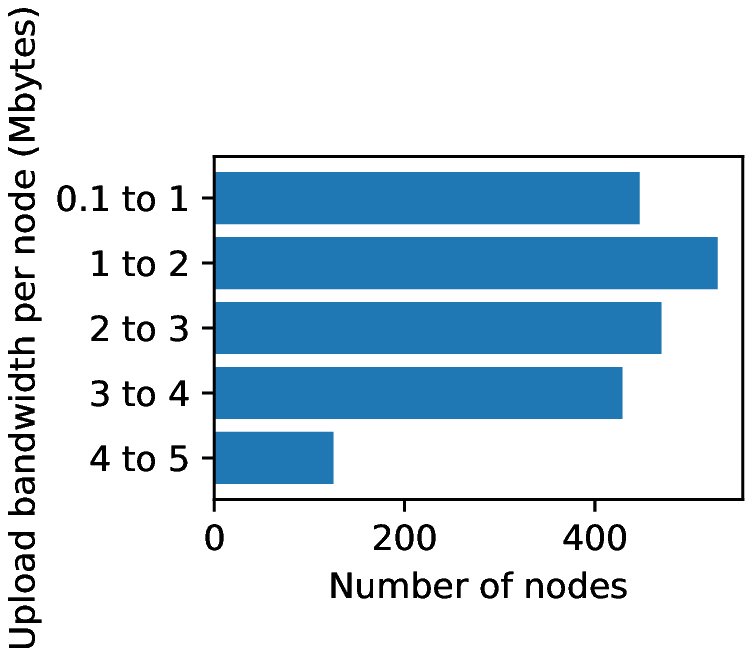} \\
             $\ $ $\ $ $\ $\footnotesize {Download bandwidth per node}&$\ $ $\ $ $\ $ \footnotesize {Upload bandwidth per node}\\
        \multicolumn{2}{c}{\includegraphics[width=0.23\textwidth]{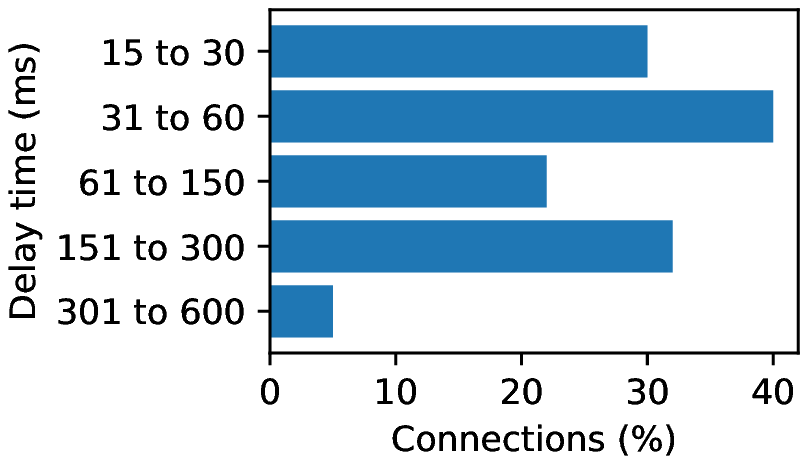}}\\
        \multicolumn{2}{c}{\footnotesize{Delay time per connection}}
    \end{tabular}
    \captionof{figure}{Basic statistic of network}
    \label{fig:img6}
\end{table}

For Bitswap network, we set a DNS server that returns information of 3 to 5 random nodes to the inquirer. In every block interval, the node asks the DNS server for new nodes and build a connection with these nodes. All nodes re-transmit data to its peers with the $P(send|r)$ possibility. After block height $4000$, we begin to compare the performance of re-transmitting of both contract-connection and Bitswap. We conducted $100$ times of tests; we randomly select nodes as the data publisher and send the data sized $1MBytes$. As the two networks are the mirror image to each other (the connections are different, but the node capacity are the same), we say every test starts from the same node. Figure \ref{fig:img8} shows a comparison between the two networks. As can be seen from the result, the broadcasting speed is mostly stable in contract-connection than in Bitswap, and the general propagation time expectation is much lower in contract-connection. It is safer to reduce the block interval in contract-connection without afraid causing centralisation. 
\begin{table}[]
    \centering
    \begin{tabular}{c}
             \includegraphics[width=0.35\textwidth]{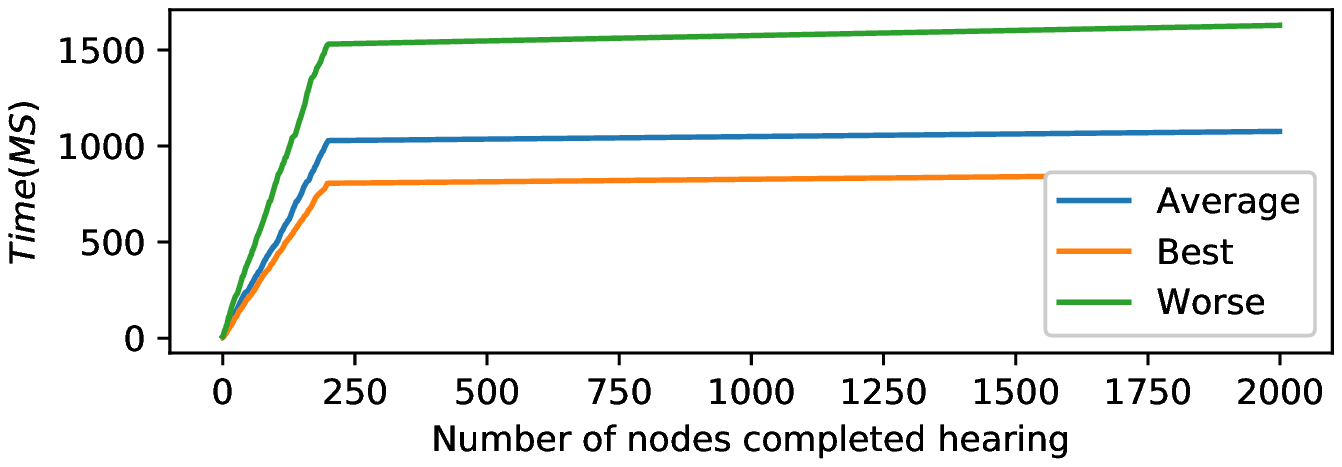}\\
            Contract-connection performance\\
        \includegraphics[width=0.35\textwidth]{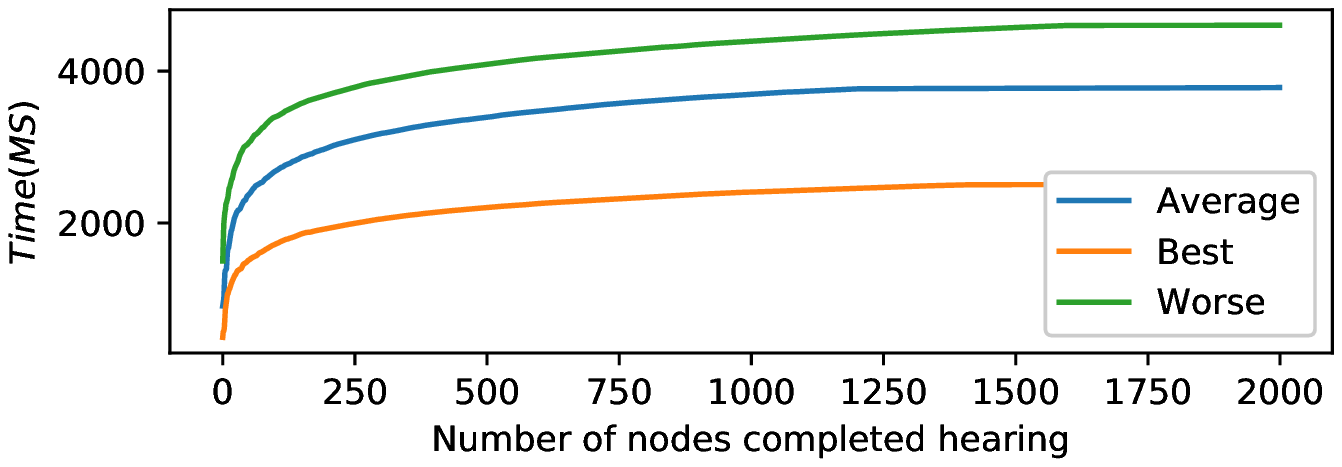}\\
            Bitswap performance\\
    \end{tabular}
    \captionof{figure}{Contract-connection performance VS Bitswap performance.}
    \label{fig:img8}
\end{table}
\section{CONCLUSION}
In this paper, we discussed a balanced communication protocol for Distributed Ledger Technology. By writing the connection information into the blockchain, the nodes derived a tamper-resisted network topology. By building a link between the node's peer structure with its general performance and setting restriction for peering, a quantified performance score for every node is periodically calculated. Through the Q-learning algorithm, every node attempts to higher its ability to hear from all the directions of the network. As the experiment suggests, the whole network is balanced during the nodes making their local optimisation.

%%%%%%%%%%%%%%%%%%%%%%%%%%%%%%%%%%%%%%%%%%%%%%%%%%%%%%%%%%%%%%%%%%%%%%%%%%%%%%%%

%%%%%%%%%%%%%%%%%%%%%%%%%%%%%%%%%%%%%%%%%%%%%%%%%%%%%%%%%%%%%%%%%%%%%%%%%%%%%%%%

%%%%%%%%%%%%%%%%%%%%%%%%%%%%%%%%%%%%%%%%%%%%%%%%%%%%%%%%%%%%%%%%%%%%%%%%%%%%%%%%

\bibliographystyle{unsrt}
{

\bibliography{sample}
}
\end{document}